\documentclass[traditabstract]{aa}
\usepackage{txfonts}
\usepackage{epsf}
\usepackage{graphicx,lineno}
\usepackage{txfonts}
\usepackage{float}
\usepackage{setspace}
\usepackage[latin1]{inputenc}   
\usepackage{graphics}
\usepackage{amssymb}
\usepackage{epsfig}
\usepackage{bibentry}
\usepackage{multirow}
\usepackage[dvips]{color}
\usepackage{natbib}
\usepackage{longtable}

\bibpunct{(}{)}{;}{a}{}{,} 

\def\apj{Astrophysical Journal}
\def\apjl{Astrophysical Journal Letters}

\def\nat{Nature}
\def\icarus{Icarus}

\def\aap{Astronomy \& Astrophysics}

\def\jgr{Journal of Geophysical Research}

\def\planss{Planetary \& Space Sciences}

\begin{document}

\title{Venus transit 2004: Illustrating the capability of exoplanet transmission spectroscopy}
\author{
P. Hedelt \inst{1,2} \and
R. Alonso \inst{3} \and
T. Brown \inst{4} \and
M. Collados Vera \inst{5} \and
H. Rauer \inst{6,7} \and
H. Schleicher \inst{8} \and
W. Schmidt \inst{8} \and
F. Schreier \inst{9} \and
R. Titz \inst{6}
}

\institute{
CNRS, UMR 5804, Laboratoire d'Astrophysique de Bordeaux, 2 rue de l'Observatoire, BP 89, F-33271 Floirac Cedex, France \and
Université de Bordeaux, Observatoire Aquitain des Sciences de l'Univers, 2 rue de l'Observatoire, BP 89, F-33271 Floirac Cedex, France \and
Observatoire Astronomique de l'Universit\'e de Gen\`eve, 51 chemin des Maillettes, 1290 Sauverny, Switzerland \and
Las Cumbres Observatory Global Telescope, 6720 Cortona Dr. Ste. 102, Goleta, CA 93117, USA \and
Instituto de Astrofisica de Canarias, 38200 La Laguna, Tenerife, Spain \and
Institut für Planetenforschung, Deutsches Zentrum für Luft- und Raumfahrt, Rutherfordstr. 2, 12489 Berlin, Germany \and
Zentrum für Astronomie und Astrophysik, Technische Universität Berlin, Hardenbergstr. 36, 10623 Berlin, Germany \and
Kiepenheuer-Institut für Sonnenphysik, Schöneckstr. 6, 79104 Freiburg, Germany \and
Institut für Methodik der Fernerkundung, Deutsches Zentrum für Luft- und Raumfahrt, Oberpfaffenhofen, 82234 Weßling, Germany
}
\date{Received ... / Accepted ...}
\abstract{The transit of Venus in 2004 offered the rare possibility to remotely sense a well-known planetary atmosphere using ground-based observations for absorption spectroscopy.  Transmission spectra of Venus' atmosphere were obtained in the near infrared using the Vacuum Tower Telescope (VTT) in Tenerife. Since the instrument was designed to measure the very bright photosphere of the Sun, extracting Venus' atmosphere was challenging. CO$_2$ absorption lines could be identified in the upper Venus atmosphere. Moreover, the relative abundance of the three most abundant CO$_2$ isotopologues could be determined. The observations resolved Venus' limb, showing Doppler-shifted absorption lines that are probably caused by high-altitude winds.

This paper illustrates the ability of ground-based measurements to examine atmospheric constituents of a terrestrial planet atmosphere which might be applied in future to terrestrial extrasolar planets.}
\titlerunning{Venus transit 2004}
\maketitle

\section{Introduction}\label{Intro}
Transits of Venus, during which Venus crosses the line of sight between Earth and the Sun, come in pairs separated by eight years over a total period of $\sim$121.5 years. The last transit occurred in 2004 whereas the next one will be observable in 2012. Apart from the inner planets of the solar system, extrasolar planets also have a certain possibility to transit in front of their central star, offering the feasibility of investigating their atmosphere during their primary transit. Up to now, 140 transiting extrasolar planets are known out of the more than 500 exoplanets that have been found in total\footnote{See J. Schneider's Extrasolar Planet Encyclopedia at http://exoplanet.eu, (status: June 2011)}. Out of this, seven transiting planets might be classified as terrestrial: CoRoT-7 b (4.80\,M$_\mathrm{Earth}$, \citealt{Leger2009}), GJ 1214 b (6.36\,M$_\mathrm{Earth}$, \citealt{Charbonneau2009}) as well as the recently detected planets Kepler-11 b,d,e,f (4.30, 6.10, 8.40, 2.30\,M$_\mathrm{Earth}$, respectively, \citealt{Lissauer2011}) and Kepler-10 b (4.54\,M$_\mathrm{Earth}$, \citealt{Batalha2011}).

During a planetary transit, the light emitted by the central star traverses the planetary atmosphere and is attenuated before it reaches the observer. Analysis of the extinction allows to characterize the planet, for example the chemical composition of the atmosphere. The apparent planetary radius at which the tangential optical depth is below unity varies at different wavelengths and can hence be used to infer the height of the atmosphere (related to pressure and atmospheric structure) and the existence of cloud layers (see e.g. \citealt{Seager2000,Brown2001}).

In the past, transits of Mercury offered the opportunity to detect absorption features within the atmosphere by ground-based spectroscopic observations. For example, during the Mercury transit in 2003, neutral sodium in its exosphere has been detected using this technique \citep{Schleicher2003}. Also the Earth can be investigated as a transiting planet during lunar eclipse observations, showing absorption features of the major atmospheric constituents like CH$_4$, H$_2$O, O$_2$, O$_3$ and CO$_2$ in reflected light from the Sun by the Moon (see e.g. \citealt{Palle2009,Vidal-Madjar2010}). Furthermore Earthshine observations, where the reflected sunlight from Earth is observed on the Moon's night side, show absorption signatures of O$_2$, O$_3$ and H$_2$O as well as a signature of Earth's vegetation red edge \citep{Arnold2002,Arnold2008}.

This technique can of course also be applied to transiting exo\-pla\-nets:
Molecular absorption bands in the infrared have been detected in the atmosphere of planets around bright stars (e.g. \citealt{Knutson2007,Tinetti2007,Swain2008}). Furthermore, the sodium resonance doublet at 598.3\,nm in the atmosphere of the transiting extrasolar giant planet HD 209458 b \citep{Charbonneau2002} during four planetary transits has been detected, as well as neutral hydrogen (HI) \citep{Vidal-Madjar2003}, oxygen (OI) and carbon (CII) \citep{Vidal-Madjar2004}. Recently, \citet{Snellen2010} used high-resolution ground-based spectroscopy to detect CO lines in the atmosphere of HD 209458 b at about 2\,300\,nm. These lines showed a blue shift of about 2\,km\,s$^{-1}$ with respect to the velocity of the host star, perhaps indicating a velocity flow from the day to the night side, also known as subsolar-to-antisolar (SSAS) flow.

However, performing transmission spectroscopy of terrestrial exoplanets and detecting the faint atmospheric absorption lines is a difficult task. Up to now, only one terrestrial exoplanet has been characterized spectroscopically using transmission spectroscopy: \citet{Bean2010} obtained transmission spectra from the terrestrial exoplanet GJ 1214b in the range from 780 to 1\,000\,nm, showing a lack of spectral features which indicates either a dense, water vapor dominated atmosphere or a hydrogen dominated atmosphere with a cloud layer which is opaque in the observed wavelength range. Additional observations by \citet{Croll2011} and \citet{Desert2011} rule out a fully water dominated atmosphere. Currently a solar composition atmosphere without methane and Rayleigh scattering by clouds best fits the data \citep{Miller-Ricci2011}.

Major modeling efforts for synthetic spectra of hypothetical terrestrial exoplanets were performed by  \citet{DesMarais2002,Segura2003,Segura2005,Tinetti2006,Ehrenreich2006,Kaltenegger2007,KalteneggerTraub2009} and \citet{Rauer2011}, in which the terrestrial planets of the solar system (i.e. Venus, Mars and Earth) are investigated as if they were distant exoplanets. These studies also include different parameter studies, in which e.g. the central star was varied, the orbital distance or the atmospheric composition. Note that \citet{Ehrenreich2006} has studied how Venus would look like if seen as a transiting exoplanet, when orbiting different main-sequence stars.

This paper investigates the feasibility of the transmission spectroscopy technique for analyzing the atmosphere of a terrestrial (exo-) planet by using ground-based high-resolution measurements during a Venus transit. A cloud-free radiative transfer model was used to compare these measurements with synthetic transmission spectra of Venus' upper atmosphere above the cloud top, assuming a pure CO$_2$ atmosphere.

In Sect. \ref{Obs} the observations and the data reduction are described and the extracted limb spectra are discussed. The radiative transfer model and the calculations are described in Sect. \ref{RadTrans}. In Sect. \ref{ParVar} a parameter variation is performed, before the final fit to measured data is performed in Sect. \ref{Fit}. Finally the results are summarized in Sect. \ref{Summary} and implications for extrasolar planets are discussed in Sect. \ref{Implications}.

\section{Observations and Reduction}\label{Obs}
The transit of Venus on June 8th, 2004 was observed at the Vacuum Tower Telescope (VTT) of the Kiepenheuer Institut für Sonnenphysik (KIS) in
Tenerife. The observations of the transit lasted from 06:30h (UT), when the Sun rose with Venus already in front of the disk, until 11:26h (UT) at fourth contact. The VTT is a 70\,cm diameter telescope with a focal length of 46\,m, equipped with an Echelle spectrograph. The spectrograph slit corresponds to a width of 0.35$\arcsec$ and a length of 92$\arcsec$ on the sky. In the focal plane of the spectrograph, the Tenerife Infrared Polarimeter (TIP, \citealt{TIP}) detector was installed. It consists of four 128$\times$128 pixel CCDs and was used in its spectrograph mode (without polarimeter).
The spectral resolution of the entire system was $ R= \lambda / \Delta \lambda =$ 200\,000. The image scale in the focal plane was 0.35$\arcsec$\,px$^{-1}$. Thus one pixel corresponds to about 71\,km at the position of Venus.

Three wavelength regions (named by the observers) with strong absorption lines of CO$_2$ were observed with a sampling of 0.031\,$\rm\AA$\,px$^{-1}$:
\begin{itemize}
\item{1\,596.49\,nm - 1\,597.29\,nm (``Favorite'') (1\,278 images)}

\item{1\,597.49\,nm - 1\,598.27\,nm (``Favorite+'') (4\,779 images)}

\item{1\,612.39\,nm - 1\,613.19\,nm (``Grabbag'') (4\,103 images)}
\end{itemize}

One single image is an integration of 20 exposures with an exposure time of 50\,ms to provide 1\,s integration time. Under the same conditions, 207 flatfield images on the unobscured solar disk and 127 darks were taken.

The telescope tracked the Sun and drift scans over the Venus disk were performed, i.e. Venus moved across the slit that
was fixed with respect to the solar image. For the observations in the ``Favorite'' wavelength range, the slit was oriented at an angle of
315.5$\degr$ (measured from Venus' north pole, anti-clockwise). During the observations in the ``Grabbag'', as well as in the ``Favorite+''
wavelength range, the slit was rotated to an angle of 45.5$\degr$ (measured from the Venusian north, anti-clockwise) and the observation was further accomplished within this position.

Note that \citet{Alonso2006} examined the data with respect to latitudinal variations of absorption line positions, observing different depths for the spectral features in the ``Favorite+'' region at higher and lower latitudes, but without further interpretation of the data. In this paper, however, an average over the different latitudes is used and the absorption of the three most abundant isotopologues of CO$_2$ is thus investigated with a higher signal-to-noise ratio (SNR).

Note that the ``Grabbag'' region finally turned out not to be usable due to a strong noise pattern and a strong gain jump between the right and left CCDs, which could not be corrected. Therefore transmission spectra of this region are omitted in the following.

\begin{figure}[h]
\center
a) \includegraphics[width=8cm]{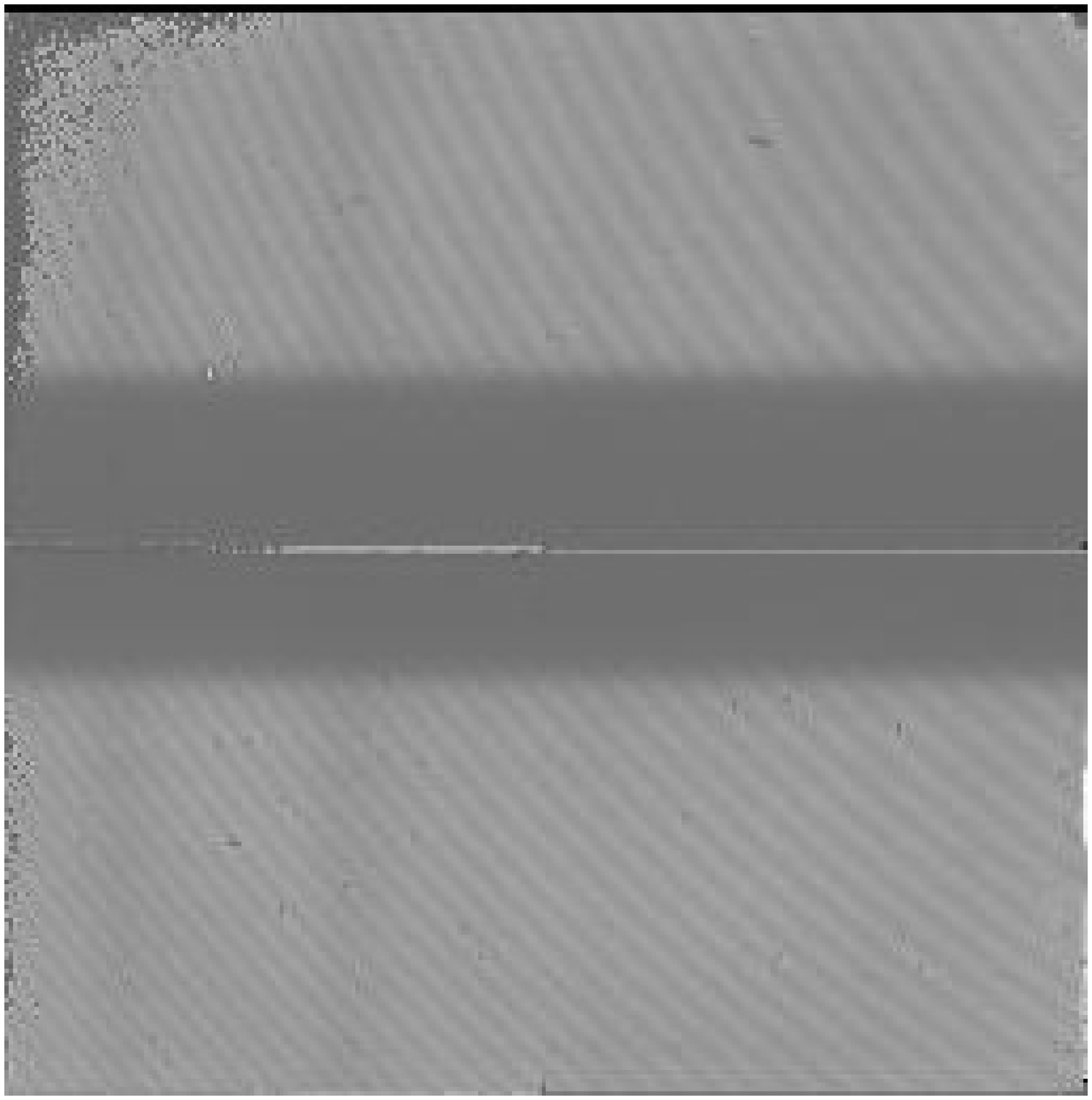}\\
b) \includegraphics*[width=8cm]{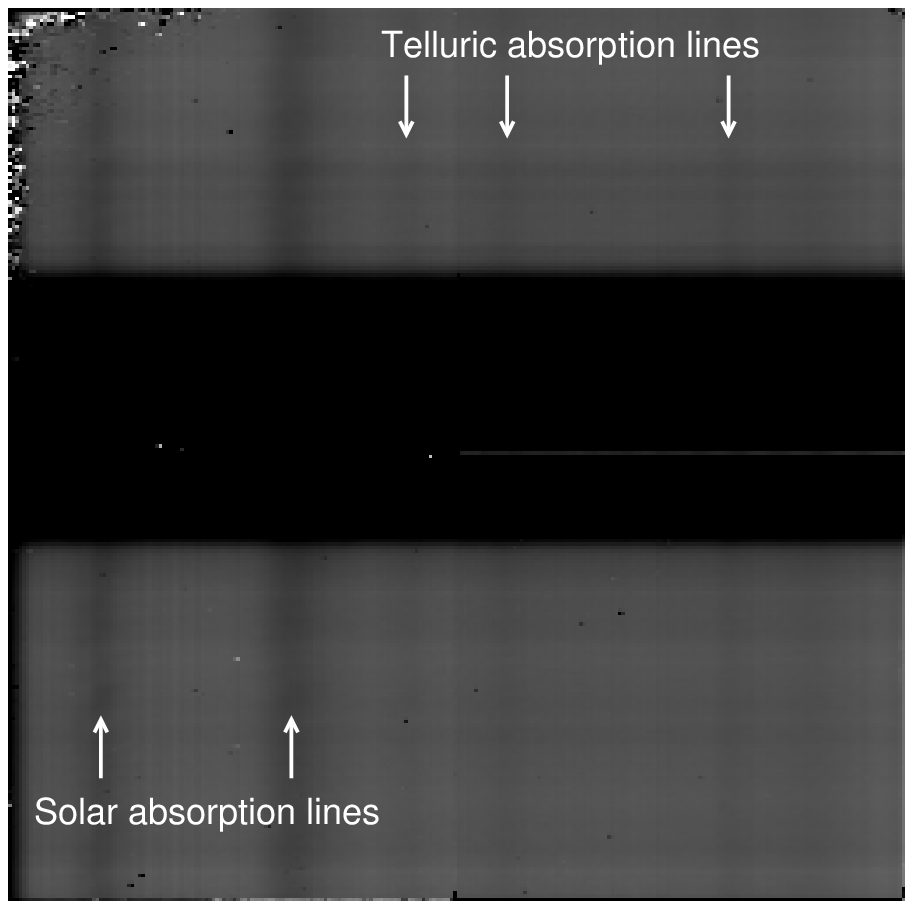}
\caption{a) Raw image from the ``Favorite+'' region: Clearly visible is the Venus disk as the horizontal dark strip in the
center of the image, the fringe pattern (sloping dark lines), the bad pixel area in the upper left corner and the gain jump between the upper and
lower CCD chips in the center of the frame. Strong solar and weak telluric absorption lines appear as vertical dark stripes. b) Reduced image, after
flatfielding. Clearly visible are the solar and telluric absorption lines and the gain jump between the CCDs.}\label{Figure1}
\end{figure}

An example of a raw image (1s integration time) is shown in Fig. \ref{Figure1}a. It shows a ring-like fringe pattern, a central horizontal line separating the two upper 128$\times$128 pixel CCDs from the lower ones, and a region with a noticeable number of bad pixels in the upper left corner.
In the figure the y-axis is the position, whereas the x-axis is the wavelength. Slightly dark vertical bands show the solar and terrestrial absorption lines. The disk of Venus can be seen as a thick horizontal band in the center of the image.

A standard image reduction procedure has been used to calibrate the images, including dark current removal and flatfielding. Note that the final flatfield master image did not contain any absorption lines of the Sun or Earth, but only the fringe pattern. Thus in the final reduced images (see Fig. \ref{Figure1}b), only the interference fringes are removed whereas solar and telluric absorption lines are now clearly visible as well as the separation between the CCDs.

The next step was to extract the spectra of Venus' atmosphere. First, the position of both Venus limbs were identified on the reduced images and taken as reference points. The limb positions were determined by fitting the peaks of the first derivative of the spatial intensity with a Gaussian.

The mean diameter (taking into account the pixel scale of the CCD) of the projected Venus disk in the ``Favorite'' region was 7\,980.018$\pm$2\,164.985\,km, whereas in the ``Favorite+'' region it was 4\,683.8790$\pm$971.466\,km. Thus the slit was never positioned over the center of the Venus disk, but more to the limb, and hence the lower limit on the projected atmospheric altitude covered by one pixel is 46.313$\pm$12.565\,km in the ``Favorite'' region and 27.184$\pm$5.638\,km in the ``Favorite+'' region, assuming that the radius of the visible disk of Venus is 6\,116.80\,km (hence including the opaque cloud deck at 65\,km altitude). Therefore the upper atmosphere can be resolved within a few pixels.

After that, a spectral and spatial normalization was performed: First, each image row was divided by a spectrum of the solar disk, which was
obtained by binning two image rows 11 pixels offset from the Venus limb. This step removes the telluric absorption lines but only attenuates the solar absorption lines. Note that in order to distinguish between telluric absorption lines and lines forming in the atmosphere of an exoplanet, ground-based observations of transiting exoplanets may take advantage of the Doppler shift due to the proper motions of the star, the exoplanet, the Sun and Earth (see e.g. \citealt{Vidal-Madjar2010}).

\begin{figure}[h]
\center
\includegraphics*[width=8cm]{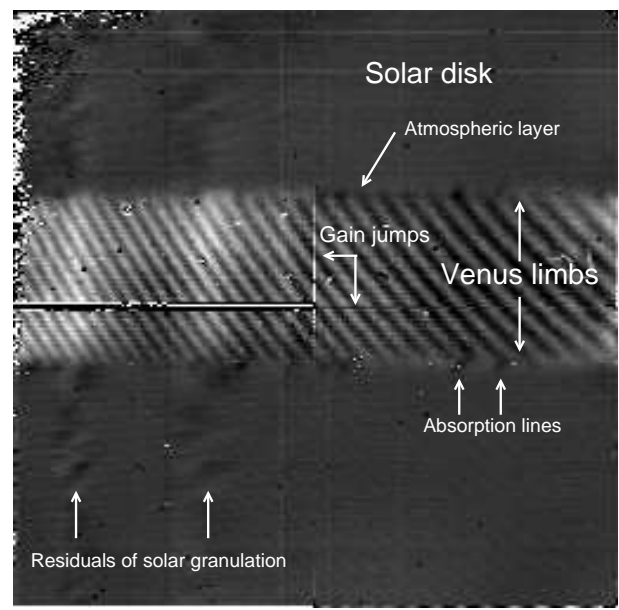}
\caption{Flatfielded and normalized image from the ``Favorite+'' region. In the transition
region between the solar surface and the disk of Venus a thin layer with absorption features is visible, which represents the atmospheric layers of Venus.}\label{Figure2}
\end{figure}

Next, the spatial normalization was performed by averaging image columns in a clean part of the CCD and dividing the image by this.  The resulting image in Fig. \ref{Figure2} now clearly shows weak absorption lines in the transition region between the Venusian and the solar disk, which do not correlate with the solar absorption lines. Comparison with the HITRAN2004 (High-Resolution Transmission Molecular Absorption Database, \citealt{Rothman2005}) line database identifies the atmospheric absorption lines as those of CO$_2$. The flatfielding did not work properly at the position of the Venus disk, as the fringe pattern is still visible.

The resulting residuals of solar absorption lines visible in Fig. \ref{Figure2} arise from Doppler shifts of convecting granules in the photosphere of the Sun and the so called ``5 minute'' oscillation. The Doppler shift of these residuals is $\sim$0.0143\,nm, which corresponds to a velocity of 1.7\,km\,s$^{-1}$ and a size of $\sim$1\,300\,km. These values are consistent with values measured by \citet{Carroll1996}.

Finally, the spectra of Venus' atmosphere at the limb position were extracted and all spectra of a particular wavelength region were averaged, in order to improve the SNR. Thus the information about the altitude and the corresponding Doppler-shifts by winds are lost. Note that the afore mentioned normalization is not perfect, primarily due to varying gain jumps between the CCDs and at wavelengths with strong solar absorption lines. This results in a distorted baseline in the averaged spectrum at each limb. The baseline in every image is therefore corrected by applying a ninth degree polynomial fit. It was checked that the relative absorption depths were not affected by this correction. Furthermore, on some images the atmospheric layers are disturbed by the fringe pattern within the Venus disk, inhibiting any detection of absorption lines. These images were removed manually before averaging all spectra.

\begin{figure}[h]
\center a)
 \includegraphics[width=8cm]{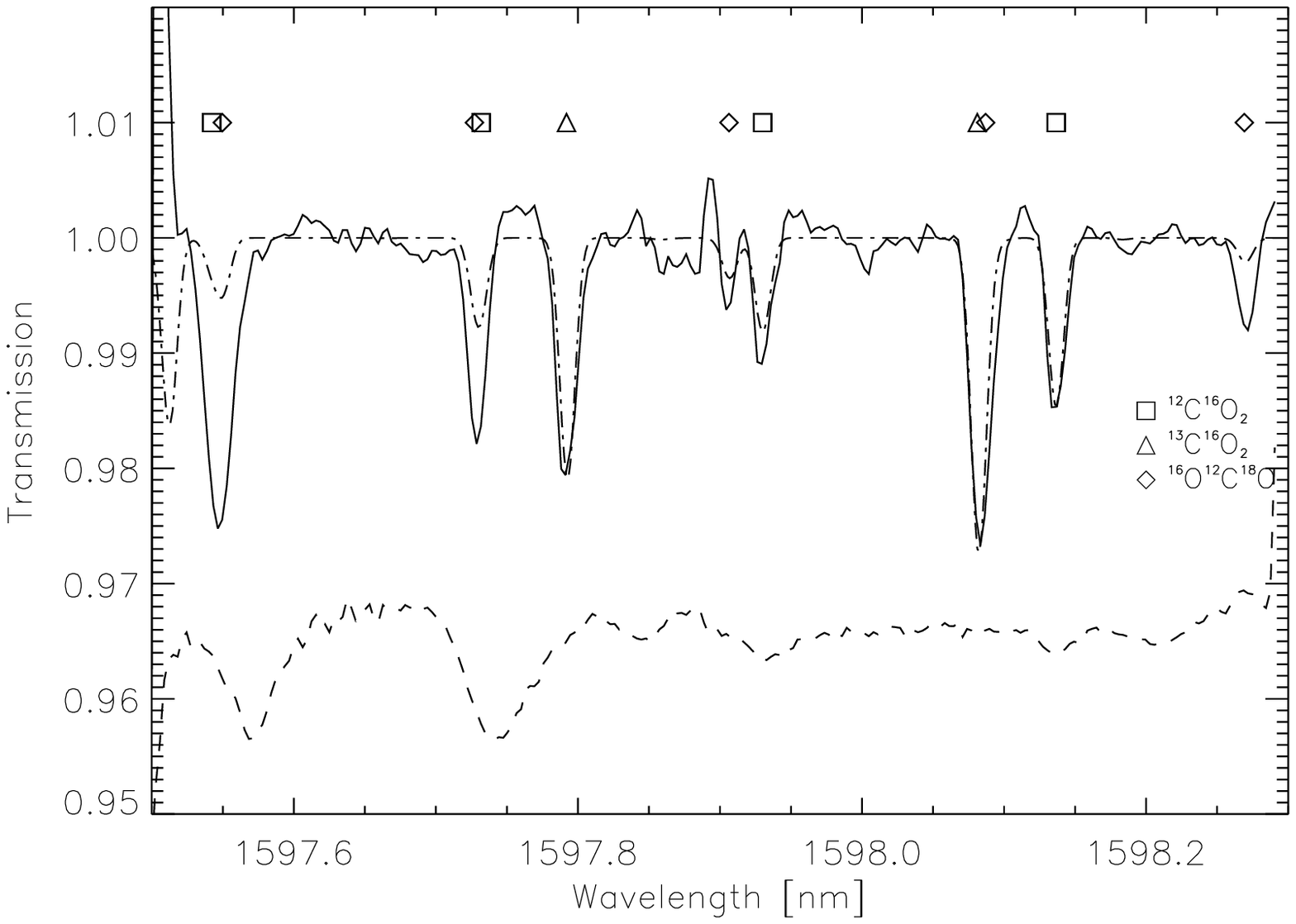}\\
b)\includegraphics[width=8cm]{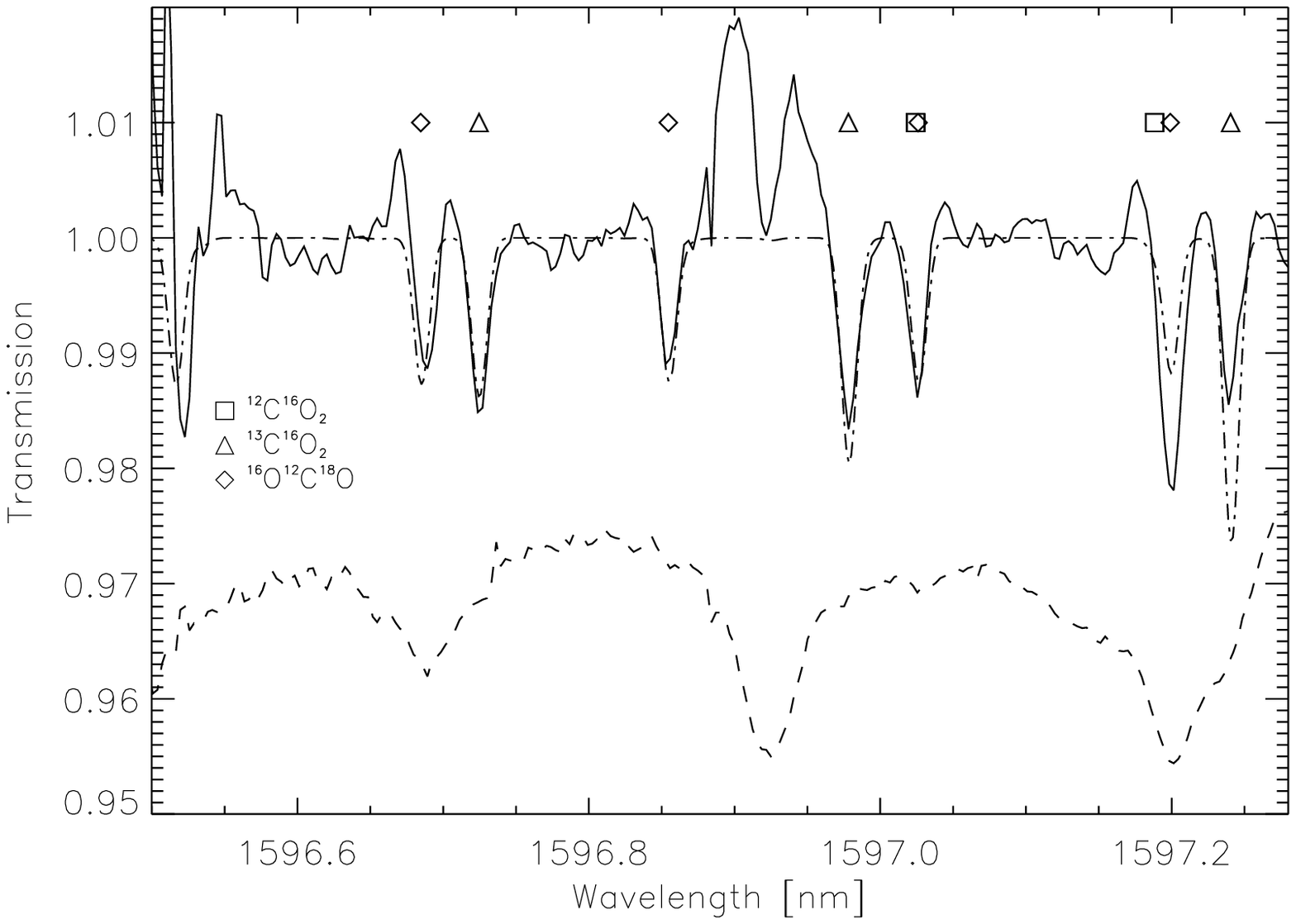}
\caption{Averaged spectrum (solid line) of both Venus limbs in the a) ``Favorite+'' and b) ``Favorite'' region together with the best fitting model spectrum (dash-dotted line, see text for details). Absorption line positions of the three isotopologues are indicated by different symbols.
\newline The dashed line shows a scaled spectrum of the Sun in arbitrary units, showing partial overlap of solar absorption lines with absorption lines from Venus' atmosphere.}\label{Figure3}
\end{figure}

In total 395 images from the ``Favorite'' and 1\,107 images from the ``Favorite+'' region have been averaged. Using line positions from HITRAN as a reference, an accurate wavelength calibration was performed. See Figs. \ref{Figure3} a) and b) for the averaged spectra from the ``Favorite+'' and ``Favorite'' wavelength range, respectively. Note that the Doppler shift of absorption lines due to the high wind velocities in the upper atmosphere of Venus was neglected, although superrotation plays a major role in the upper atmosphere dynamics.

However, still a Doppler shift between adjacent pixel rows can be measured: Note that the Venus absorption lines in Fig. \ref{Figure2} are tilted on both sides of the limb, with a blue shift of the lines with increasing altitude on the eastern Venus limb (lower part of the Venus disk in Fig. \ref{Figure2}), and a red shift of the absorption lines on the western limb. The Doppler shifted lines yield a velocity gradient of about 7\,m\,s$^{-1}$ per km altitude, which remains nearly constant with increasing altitude on the western limb, while it decreases slightly on the eastern limb.

This might be caused by a combined effect of the superrotation of Venus' mesosphere and the SSAS flow in the upper atmosphere \citep{Bougher1997}. SSAS wind velocities measured by \citet{Goldstein1991} and \citet{Betz1977} at an altitude of 100\,km are in the range from 105 to 140\,m\,s$^{-1}$, whereas recent measurements by \citet{Lellouch2008} from the Venus Express mission show wind speeds in the range from 30 to 50\,m\,s$^{-1}$ at $\sim$93\,km altitude strongly increasing up to 90 to 120\,m\,s$^{-1}$ at $\sim$102\,km altitude. The velocity gradient measured from the Venus transit spectra is thus in agreement with the gradient determined by \citet{Lellouch2008}. A detailed study of this effect is however beyond the scope of this paper and will be the subject of a following paper, perhaps incorporating observations during the next Venus transit in June, 2012.

\section{Radiative transfer calculations} \label{RadTrans}
The radiative transfer model SQuIRRL (Schwarzschild Quadrature InfraRed Radiation Line-by-line, \citealt{Schreier2001}) has been used to fit the measured spectra. SQuIRRL is a cloud free, line-by-line radiative transfer model which uses the molecular absorption line database HITRAN for the calculation of absorption cross sections. SQuIRRL calculates emission and absorption spectra for arbitrary atmospheric paths, assuming local thermodynamic equilibrium (LTE). Furthermore a convolution with appropriate spatial and spectral instrumental response functions can be performed.

The model atmosphere extends from 0 to 180\,km altitude, divided into 30 layers. CO$_2$ is treated as the only atmospheric absorber with an isoprofile of 96.5\% abundance up to 100\,km altitude. Above, the CO$_2$ profile from \citet{Hedin1983} was used. For the transmission calculations, below 100\,km altitude the day side temperature profile from \citet{Seiff1985} was used, whereas above the profile from \citet{Hedin1983} was adopted, which is in agreement with recent solar occultation measurements of Venus' mesosphere from the Venus Express mission (see e.g. \citealt{Mahieux2010}).

Transmission spectra for a set of different tangential heights above 100\,km altitude are calculated for each wavelength region, using a Gaussian beam of 24\,km and 15\,km half-width half-maximum (HWHM) for the ``Favorite'' and ``Favorite+'' range, respectively, according to the average projected atmospheric height in the respective wavelength region. Note that although one pixel corresponds to about 71\,km at the position of Venus as mentioned in Sect. \ref{Obs}, the slit was not positioned perpendicular to the surface, but under a certain angle since the spectra were taken close to the limb of the Venus' disk. Thus the projected atmospheric height is used in order to account for the field of view. The resulting spectra are finally convolved with a Gaussian of $6.15\times 10^{-3}$\,nm HWHM to account for the spectral resolution of the instrument.

Note that in the radiative transfer model no clouds or aerosol opacity is included. Nevertheless, the measured spectra can be fitted quite well since the best fitting tangential heights of the extracted spectra are well above an altitude of 100\,km, as will be shown later. Above 100\,km, clouds can be neglected to a good approximation since extinction by  H$_2$SO$_4$ cloud particles is expected to extend only up to about 80\,km altitude (see e.g. \citealt{Esposito1983,Roos1993} or \citealt{Grinspoon1993}). Venus Express SPICAV/SOIR observations by \citet{Wilquet2009} have shown that the extinction coefficient of the atmospheric haze measured at 1\,553.7\,nm decreases by about two orders of magnitude to about $2\times10^{-4}$\,km$^{-1}$ at 90\,km, which is by far lower than the CO$_2$ absorption coefficient and thus can be neglected for the altitudes probed in this paper.

Furthermore, the assumption of LTE for the radiative transfer calculations may not hold in the altitude range considered in this paper. \citet{Roldan2000,Lopez-Valverde2007} have found a departure from LTE conditions above 90\,km altitude (depending on the CO$_2$ band), producing partly strong emission lines. Nevertheless, non-LTE conditions in the wings of the 1\,600\,nm absorption band, as considered in this paper, can be neglected due to several reasons:
\begin{itemize}
\item{Under non-LTE conditions, the excitation temperature of the CO$_2$ (30011, 30012, 30013) levels (which correspond to the observed absorption lines) by solar excitation is about 400\,K at altitudes above about 120\,km in the wavelength range considered here (M. L\'{o}pez-Puertas, priv. comm.). This temperature is higher than the atmospheric temperature of about 300\,K under LTE conditions (see Fig. \ref{Figure4}). However, the solar flux with a blackbody temperature of about 5\,800\,K is much higher than the flux emitted in the Venus atmosphere. }

\item{\citet{Lopez-Valverde2007} found only a very weak atmospheric emission (which is below the detection limit of the VIRTIS instrument on Venus Express) for the 1\,600\,nm CO$_2$ band at the limb for low solar zenith angles (i.e. for the transmission geometry considered in this paper) which is even lower in the band wings where the Venus transit observations have been performed.}
\item{Potential non-LTE effects caused by the non-LTE populations of the lower vibrational level of hot bands can also be neglected, since the observed lines correspond to a fundamental transition from the ground vibrational level.}
    \end{itemize}

To summarize, the contribution of the non-LTE atmospheric emission is in any case much smaller than the Sun's radiation and thus strong lines seen in emission are not expected in the considered wavelength range. Also an influence on the absorption can be neglected. Note that observations of the next Venus transit in 2012 may provide a much better data quality and may be achieved at wavelengths where non-LTE effects can not be neglected. In that case a more detailed radiative transfer code needs to be used, which includes non-LTE effects as well as cloud and aerosol extinction.

Note that for a transiting exoplanet the temperature profile is not available. The day side temperature profile might be inferred from atmospheric modeling and secondary eclipse observations by investigating the emission spectrum. However, for transmission observations the terminator temperature profile cannot be inferred easily, since it depends on numerous factors, like planetary rotation, atmospheric composition and dynamics, distance to star, etc.. Nevertheless, transmission observations of the Hot-Jupiter planet HD 209458 b by \citet{Sing2008} and \citet{Vidal-Madjar2011} were able to infer a temperature profile at the terminator by investigating sodium absorption line profiles.
Close-in exoplanets which may have a thick atmosphere and rotate slowly will be more likely to have a strong SSAS flow in order to transport warm air from their hot day side to the cold night side, thus a day side temperature profile might be applicable for the terminator region. Furthermore these atmospheres will have a higher atmospheric scale height on the day side (see for example \citealt{Burrows2010}, showing this effect on a transiting giant exoplanet), which could in principle be detected from determining the radius during the ingress and egress. However, this effect is expected to be quite small.

\section{Parameter variation}\label{ParVar}
In order to determine the best fitting model parameters, a parameter variation was performed to investigate the temperature and height dependence of the absorption lines. Furthermore the abundances of the three isotopologues found in the data was varied in order to infer isotopic ratios for $^{12}$C/$^{13}$C and $^{16}$O/$^{18}$O in Venus' atmosphere.

\begin{figure}[h]
\center \hspace*{-0.7cm}\includegraphics[width=9cm]{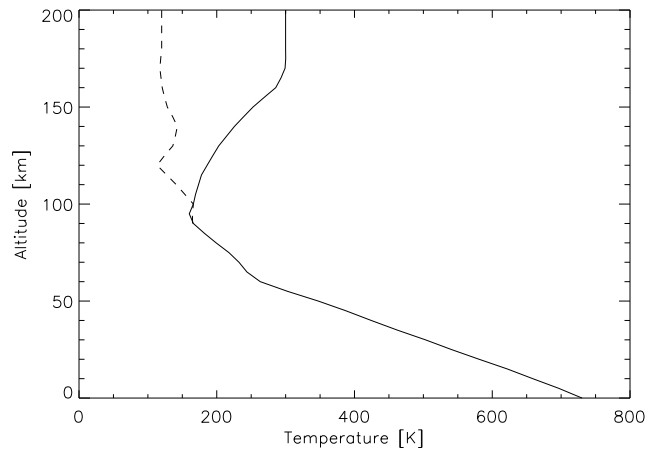} \caption{Temperature profiles used for the calculation of transmission spectra, obtained from \cite{Yung1999}. \textsl{Solid line}: Day side, \textsl{Dotted line}: Night side.}\label{Figure4}
\end{figure}

\begin{figure}[h]
\center \hspace*{-0.7cm}\includegraphics[width=9cm]{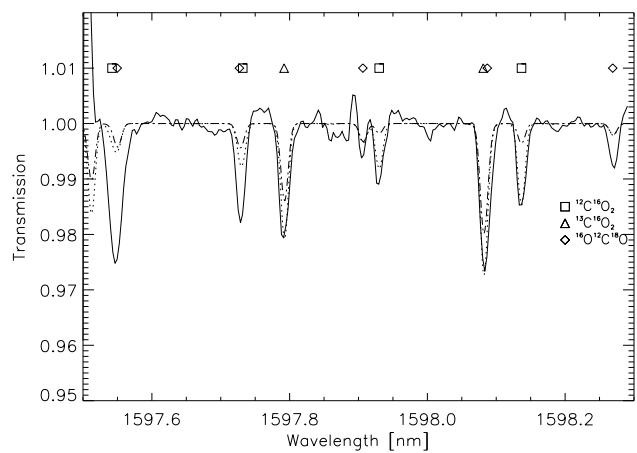} \caption{Spectral response when varying atmospheric temperature profile. \textsl{Solid line}: Measured spectrum from the ``Favorite+'' region, \textsl{Dotted line}: Spectrum calculated with the day side temperatures profile and  \textsl{Dash dotted line}: Spectrum calculated with the night side temperature profile. Absorption line positions of the three isotopologues are indicated by different symbols.}\label{Figure5}
\end{figure}

Although a fixed temperature profile was used for the final fit, the profile was varied within the temperatures of the day and night side in order to quantify the response of the absorption. Especially in the upper atmospheric layers above 100\,km, large differences in temperature between the day and night side of Venus occur (see Fig. \ref{Figure4}), reaching about 150\,K above altitudes of 170\,km. Increasing the temperature from the night to the day side profile, the absorption of $^{12}\mathrm{C}\,^{16} \mathrm{O}_2$ and $^{13}\mathrm{C}\,^{16} \mathrm{O}_2$ increases by about 0.5\% as can be seen in Fig. \ref{Figure5}. The temperature variation has no significant effect on $^{12}\mathrm{C}\,^{16}\mathrm{O}\,^{18}\mathrm{O}$.

The tangential height of the beam was varied in steps of 1\,km in the altitude range from 100 to 200\,km. With an increase of 1\,km in tangential height, the absorption of all lines decreases by about 1.3\%. Since $^{12}\mathrm{C}\,^{16}\mathrm{O}_2$ is the most abundant isotopologue and hence independent of the isotopic ratio, lines of this isotopologue are used to determine the best fitting tangential height of the measurements. Note that the determination of the tangential height is of course also dependent on the chosen beam width.

Using initially Venus isotopic ratios of $^{12}$C/$^{13}$C and $^{16}$O/$^{18}$O (see also Table \ref{Table1}), isotopic ratios were varied by large amounts, i.e. in the range from half up to double isotopic ratios of Venus. The abundances of the CO$_2$ isotopologues can be calculated according to the relationship from \citet{Bezard1987}:
 \begin{eqnarray}\label{IsoRat}
\nonumber^{12}\mathrm{C}/^{13}\mathrm{C} &=& ^{12}\mathrm{CO}_2/^{13}\mathrm{CO}_2\\
^{16}\mathrm{O}/^{18}\mathrm{O} &=& 2\,^{12}\mathrm{C}\,^{16}\mathrm{O}_2/^{12}\mathrm{C}\,^{16}\mathrm{O}\,^{18}\mathrm{O}.
 \end{eqnarray}
It can be clearly seen that increasing the isotopic ratios, the abundance of $^{13}\mathrm{C}\,^{16}\mathrm{O}_2$ and $^{12}\mathrm{C}\,^{16}\mathrm{O}\,^{18}\mathrm{O}$ decrease.

\begin{table}
  \begin{center}
  \caption{Carbon and oxygen isotopic ratios}\label{Table1}

  \begin{tabular}{lccc}
    \hline\hline
    Planet & $^{12}\mathrm{C}/^{13}\mathrm{C}$ & $^{16}\mathrm{O}/^{18}\mathrm{O}$ & Reference\\\hline
    Earth & 89.01$\pm$0.38 & 498.71$\pm$0.25 & (1)\\
    Venus & 86$\pm$12 & 500$\pm$80 &(2)\\
    Venus & 110$\pm$30 & 537$\pm$53 & (3)\\\hline
  \end{tabular}
    \end{center}
   (1) \citealt{Yung1999}; (2) \citealt{Bezard1987}; (3) This work
\end{table}

\section{Spectral fitting}\label{Fit}
The fit of calculated and measured data was performed visually since it was not possible to use a least-squares fit due to the poor spectral normalization and the remaining solar residuals and noise features. Only isolated lines with a good baseline were used for the fitting which do not overlap with solar absorption lines.

\subsection{``Favorite+'' wavelength range}
In the ``Favorite+'' region (Fig. \ref{Figure3}a), the $^{12}\mathrm{C}\,^{16}\mathrm{O}_2$ absorption lines at 1\,597.93\,nm and at 1\,598.14\,nm (box symbols in the figure) can be used for the tangential height determination. The absorption strength may however be disturbed by the higher baseline nearby. The best fit using these lines was achieved with a tangential height of 116$\pm$1\,km using a beam width of 15\,km HWHM. Taking into account the uncertainty in the beam width (see section \ref{Obs}) the best fitting tangential height is 116$\pm$5\,km. All other lines of this isotopologue overlap with other isotopologues or are at positions where solar absorption lines are located. Note that the inferred tangential height is close to the homopause located at about 130\,km altitude, above which turbulent mixing of the atmosphere becomes unimportant and the atmosphere becomes stratified. Although this could influence the isotopic ratio over the altitude region probed, the atmospheric levels above the inferred tangential height do not contribute much to the absorption lines and this effect is neglected. Furthermore the data quality is too low as to resolve this effect.

The absorption lines of $^{13}\mathrm{C}\,^{16}\mathrm{O}_2$ at 1\,597.79\,nm and at 1\,598.08\,nm (triangle symbols in the figure) are used for the determination of the carbon isotopic ratio, giving a best fit of $^{12}$C/$^{13}$C=110$\pm$30, which is higher than the value for Venus inferred from \citet{Bezard1987}. However taking into account the deviated baseline, the values are within the error margin comparable to literature values. Note that although the line at 1\,598.08\,nm overlaps with $^{12}\mathrm{C}\,^{16}\mathrm{O}\,^{18}\mathrm{O}$, the absorption is dominated by $^{13}\mathrm{C}\,^{16}\mathrm{O}_2$.

The isolated lines of $^{12}\mathrm{C}\,^{16}\mathrm{O}\,^{18}\mathrm{O}$ at 1\,597.91\,nm and 1\,598.27\,nm (diamond symbols in the figure) could not be used for determining the isotopic ratio of $^{16}$O/$^{18}$O due to the spike at about 1\,597.89\,nm. Nevertheless varying the $^{16}$O/$^{18}$O ratio has only a marginal effect in the absorption. Therefore, the isotopic ratio found in the ``Favorite'' region (see hereafter) will be used here as well, with $^{16}$O/$^{18}$O=537$\pm$53.

Figure \ref{Figure3}a shows the measured spectrum together with the best fitting model spectrum with a tangential height of 116\,km, and isotopic ratios of $^{12}$C/$^{13}$C=110 and $^{16}$O/$^{18}$O=537. Additionally, a scaled solar spectrum which has been extracted from one flatfielded image of the ``Favorite+'' region is plotted (dashed) to visualize the location of the solar absorption lines.

\subsection{``Favorite'' wavelength range}
In the ``Favorite'' region (Fig. \ref{Figure3}b) the lines of $^{12}\mathrm{C}\,^{16}\mathrm{O}_2$ are very weak and masked by absorption lines of the other isotopologues and no isolated absorption lines can be found. In order to determine the best fitting tangential height, absorption lines of $^{13}\mathrm{C}\,^{16}\mathrm{O}_2$ are used, assuming that the isotopic ratio of $^{12}$C/$^{13}$C is the same as in the ``Favorite+'' region. Absorption lines at 1\,596.73\,nm and 1\,596.98\,nm provide a best fitting tangential height of 120$\pm$2\,km, when using a beam width of 24\,km HWHM. Taking into account the uncertainty in the beam width, the estimated tangential height is 120$\pm$10\,km.

In order to determine the $^{16}$O/$^{18}$O ratio, absorption lines of $^{12}\mathrm{C}\,^{16}\mathrm{O}\,^{18}\mathrm{O}$ at 1\,596.69\,nm, 1\,596.85\,nm and 1\,597.03\,nm have been chosen. The first two lines are best fitted with $^{16}$O/$^{18}$O=575, whereas the other line is best fitted with $^{16}$O/$^{18}$O=500. Note that the line at 1\,596.69\,nm may be influenced by the strong baseline deviation shortward of this line and the overlap with a solar line. Therefore the estimated best fitting isotopic ratio is $^{16}$O/$^{18}$O=537$\pm$53. This is also higher than the value determined by \citet{Bezard1987}, but still in the range of their uncertainty.

Figure \ref{Figure3}b shows the measured spectrum together with the best fitting model spectrum with a tangential height of 120\,km, and isotopic ratios of $^{12}$C/$^{13}$C=110 and $^{16}$O/$^{18}$O=537.
Like in the ``Favorite+'' region, a scaled solar spectrum is plotted (dashed) to visualize the location of the solar absorption lines.

\section{Summary and conclusion}\label{Summary}
Using the Tenerife Infrared Polarimeter at the Vacuum Tower Telescope (VTT) in Tenerife, molecular absorption lines of the three most abundant
CO$_2$ isotopologues $^{12}\mathrm{C}\,^{16} \mathrm{O}_2$, $^{13}\mathrm{C}\,^{16} \mathrm{O}_2$ and $^{12}\mathrm{C}\,^{16}\mathrm{O}\,^{18}\mathrm{O}$ have been detected within the atmosphere of Venus during its transit in June 2004. Although achieving limb spectra from the measured images was
challenging, the spectra could be modeled using a line-by-line radiative transfer model, which even allowed to determine the isotopic ratios of $^{12}$C/$^{13}$C and $^{16}$O/$^{18}$O, which are slightly higher, but still in agreement with literature values. Note that the determination of isotopic ratios shall not serve as an additional measurement to what is already known about Venus, but rather as a proof of concept in order to show, what might be possible for exoplanets in the future. By furthermore varying the tangential height while leaving the temperature profile fixed, good agreement of modeled and measured spectra could be achieved with tangential heights in the range from (116$\pm$5)\,km to (120$\pm$10)\,km, depending on the wavelength range considered.

The observations resolved Venus' limb, showing an increasing Doppler-shift of the absorption lines with increasing altitude. This is probably caused by high altitude winds in Venus' atmosphere, with a wind velocity gradient being consistent with Venus Express measurements of \citet{Lellouch2008}.

The results illustrate the feasibility of the transmission spectroscopy technique to identify atmospheric constituents and even infer isotopic ratios in terrestrial exoplanets. Future planned space missions, which are designed to measure the atmospheres of terrestrial exoplanets, will utilize similar techniques to examine their atmospheres in detail.

\section{Implications for extrasolar planets}\label{Implications}
Current models for exoplanets mostly rely on knowledge derived from the solar system planets. It is straightforward (at least conceptually) to build a spectrum from a given planet type. The inverse problem, however i.e. to infer the characteristics of a planet from a spectrum is much more difficult due to its ill-posed nature.

Transiting exoplanets cannot be resolved in front of their central star as was the case for the Venus transit presented here. Thus the method to obtain atmospheric transmission spectra of an exoplanet is different than it was performed in this work. For transits of exoplanets, where the planet and its central star are convolved within a few pixels, the total flux of the system has to be compared during and outside the transit. Since the apparent radius of the planet and thus the transit depth of the transiting planet is dependent on the wavelength due to the absorbing atmosphere, a transmission spectrum can be obtained by observing the transit at different wavelengths.

Current and near-future space telescopes provide only low resolution spectroscopy (Spitzer: R$<$600, James Webb Space Telescope: R$<$3000), thus only molecular absorption bands can be investigated rather than single absorption lines. Although ground-based telescopes already feature high-resolution spectrographs of $R\sim$100\,000 like the CRyogenic high-resolution Infrared Echelle Spectrograph (CRIRES, \citealt{Kaeufl2004}) at the Very Large Telescope (VLT), the SNR is still too low to detect lines of different isotopologues as was performed here, since atmospheric absorption lines of exoplanets are contrasted against the overall stellar flux. Atmospheric features of exoplanets are in the order of magnitude of around 10$^{-6}$, which is much smaller than the 1 to 2\% absorption detected for Venus here. This makes the investigation of the atmosphere much more challenging. Furthermore, ground-based observations have to remove telluric absorption lines in order to access the atmospheric absorption of an extrasolar planet.

Nevertheless, the investigation of an terrestrial exoplanet is already possible, as was shown for GJ 1214b \citep{Bean2010}. The data can however be interpreted in several ways and the actual composition of the atmosphere is still subject to discussion (see e.g. \citealt{Croll2011,Desert2011,Miller-Ricci2011}).

Venus composition and $p, T$ profiles are well known from satellite and ground-based observations, which is of course not the case for exoplanets. Temperature profiles of exoplanets might be inferred by emission spectroscopy during secondary eclipse observations. For example, CO$_2$ absorption bands seen in emission would indicate a temperature inversion in the stratosphere since CO$_2$ can be assumed to be well mixed throughout the atmosphere, as is the case for Venus, Mars and Earth. Also the surface temperature might be inferred in spectral windows where the atmosphere is transparent down to the ground. The thermal emission of about 30 exoplanets has already been detected, which allow the determination of the temperature structure, energy redistribution and day-night contrasts (see e.g. \citealt{Deming2005,Harrington2006,Knutson2007}).

The determination of exoplanet isotopic ratios (as shown in this paper for Venus) would give insight into the evolution and diversity of exoplanet systems. However this is beyond the ability of current and near future observational efforts, because both a significant SNR and a high spectral resolution are required to detect and separate the lines:

The Venus transit was observed with a spectral resolution of $R=$200\,000 in a wavelength range, where the three most abundant isotopologues are detectable. In the wavelength range observed, a resolution of more than about 25\,000 is required in order to separate lines of $^{12}\mathrm{C}\,^{16}\mathrm{O}_2$ and $^{13}\mathrm{C}\,^{16}\mathrm{O}_2$ to determine the $^{12}$C/$^{13}$C ratio. For the detection of $^{12}\mathrm{C}\,^{16}\mathrm{O}\,^{18}\mathrm{O}$ absorption lines for the determination of the $^{16}$O/$^{18}$O ratio, a much higher spectral resolution of about $R=$50\,000 is required, to separate them from the $^{12}\mathrm{C}\,^{16}\mathrm{O}_2$ and $^{13}\mathrm{C}\,^{16}\mathrm{O}_2$ absorption lines. However, note that other molecules might feature stronger absorption lines that are well separated.

\citet{Rauer2011} have shown for the transmission spectrum of an Earth-like planet around a solar like G star at a distance of 10\,pc that even with a low spectral resolution of $R=$2\,000 the photon-limited SNR is about 0.054 for the 4\,200\,nm CO$_2$ band, when using e.g. the James Webb Space Telescope (JWST) having an aperture of 6.5\,m. At $R=$10\,000 the corresponding SNR would be 0.022 and at $R=$50\,000 it would be 0.009, hence hundreds of transits need to be co-added in order to achieve a reasonable SNR of more than three. Since the SNR is directly proportional to the telescope aperture, a larger telescope aperture would increase the SNR. However, only with next generation telescopes having much bigger apertures of about 40\,m (e.g. the European Extremely Large Telescope, E-ELT), SNRs close to unity are achievable for terrestrial exoplanets. Observations of close-in hot-Jupiter planets will also significantly increase the SNR. Thus the determination of isotopic ratios of an exoplanet atmosphere is expected in the near future.

Despite the difficulties for high-resolution spectroscopy of extrasolar planets, the observations of Venus shown here illustrate the potential of such analysis. Once sufficiently sensitive instrumentation becomes available, the search for isotopic lines in exoplanet atmospheres can provide severe constraints on the atmospheric evolution of these distant planets.

\begin{acknowledgements}
We thank two anonymous referees for a thorough reading of and useful comments on the manuscript. We are grateful to Manuel L{\'o}pez-Puertas for comments and discussion about possible non-LTE effects in the atmosphere of Venus. P.H. acknowledges support from the European Research Council (Starting Grant 209622: E$_3$ARTHs). The VTT is operated by the Kiepenheuer Institut für Sonnenphysik at the Spanish Observatorio del Teide of the Instituto de Astrofísica de Canarias.
\end{acknowledgements}

\bibliographystyle{aa}

\end{document}